\documentstyle[editedvolume,psfig]{crckapb} 

\begin{opening}
\title{
SPECTROSCOPY OF GIANTS OF THE SAGITTARIUS \protect\\ DWARF GALAXY
       }
\subtitle{}

\author{
P. Bonifacio}
\institute{Osservatorio Astronomico di Trieste\\
           Via G.B. Tiepolo 11 I-34131 Trieste}
\author{L. Pasquini}
\institute{European Southern Observatory\\
K. Schwarzschild Strasse 2 D-85748 Garching bei Munchen - Germany
}
\author{P. Molaro}
\institute{Osservatorio Astronomico di Trieste\\
           Via G.B. Tiepolo 11 I-34131 Trieste}
\author{
G. Marconi}
\institute{Osservatorio Astronomico di Roma\\
Via dell'Osservatorio 2, Monte Porzio Catone, I-00040 Roma, Italy
}
\end{opening}

\runningtitle{Sagittarius dwarf galaxy}

\begin{document}

% The \begin{document} command comes after the \end{opening}
% command.

\begin{abstract}
In this paper we present the first results of the
analysis of intermediate resolution
($\Delta\lambda \sim 3.5 \AA$)
spectra of giants of the Sagittarius dwarf galaxy
acquired using the ESO NTT telescope.
From the deep CCD
photometry of Marconi et al 
(1998a) we have selected a sample of 
giants representative of the metallicity
spread suggested by the comparison of the colour-magnitude
diagram of Sagittarius with those of galactic globular
clusters.
The spectra have been used to measure radial velocities, to
confirm the membership to 
 Sagittarius, and to provide a 
metallicity estimate  by
using spectral synthesis techniques. 
The analyzed stars show a spread in metallicities in the range 
$-1.0\le\rm [Fe/H]\le+0.7$, some 0.5 dex more metal-rich
than the photometric estimates.

\end{abstract}

\section{Introduction}

The Sagittarius dwarf galaxy, discovered by Ibata et al (1994, 1995),
at a distance of only 25 Kpc, offers a unique opportunity to study
the stellar populations of an external galaxy.
Comparison of the color-magnitude diagram of Sagittarius with 
those of Galactic globular clusters has suggested a spread in
metallicity
in the range $-0.7
\le$ [Fe/H] $\le-1.6$ (Marconi et al 1998a).
This may be the sign of a complex star-formation history characterized
by several bursts.
Because Sagittarius is some 16 Kpc behind the Galactic Center
there is confusion between Sagittarius and Bulge stars.
Sagittarius revealed itself as a population with a mean radial 
heliocentric
velocity
around 140 kms$^{-1}$ and a small velocity dispersion, thus
probable membership may be ascribed on the basis of the radial
velocity.
Intermediate resolution spectra 
allow to determine radial velocities with a precision around 20
kms$^{-1}$, sufficient to confirm membership.
Moreover such low resolution spectra may be used to obtain a crude
estimate of the metallicity which may be compared with the 
estimates based on the colour-magnitude diagrams.
In this paper we report on radial velocities and abundances
derived from grism spectra obtained with NTT+EMMI/MOS at ESO La Silla.  

\begin{figure}
\psfig{figure=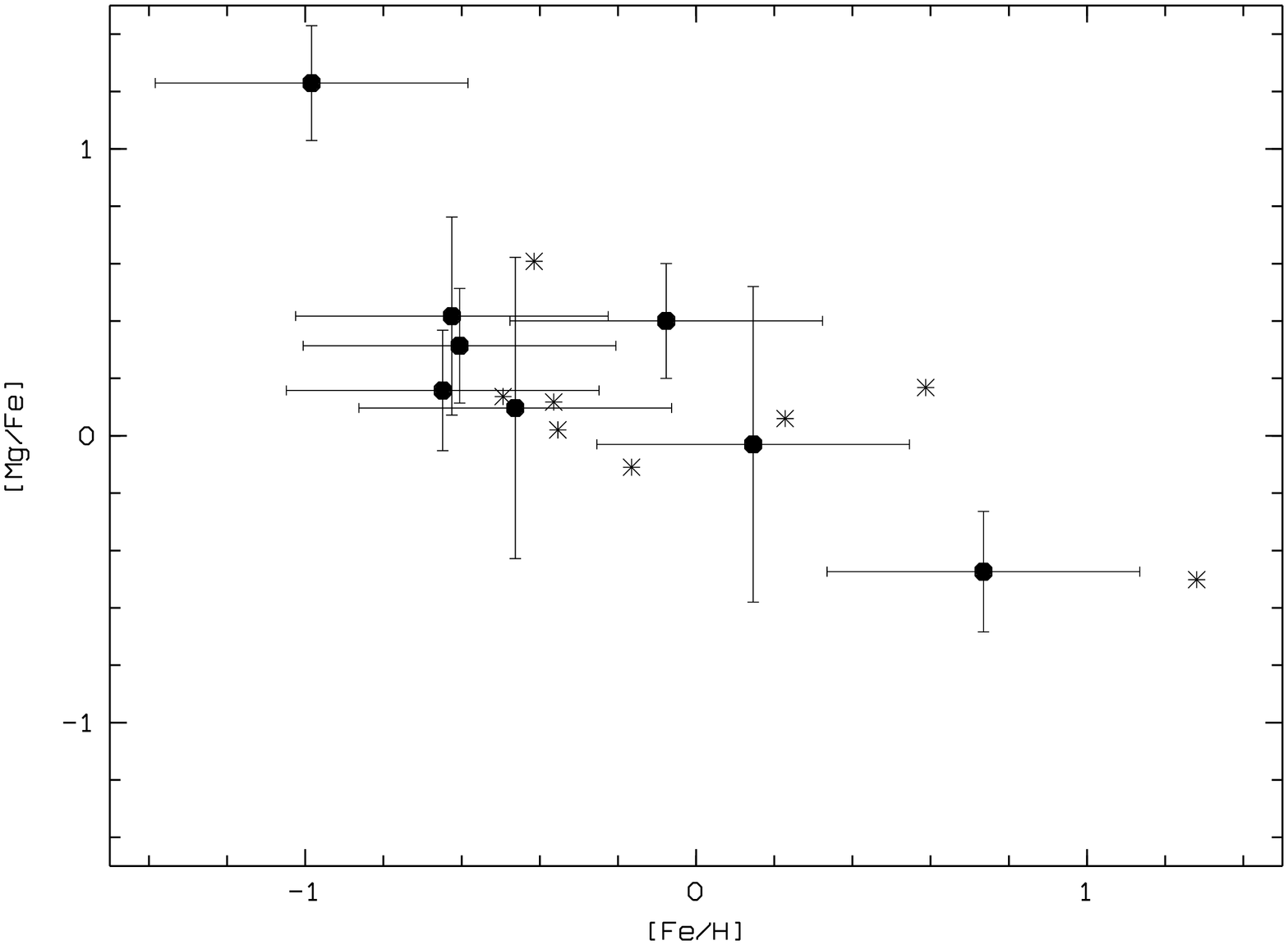,width=14cm,angle=-90}
\caption{[Mg/Fe] ratios 
versus [Fe/H] for $\xi = 2.0$ kms$^{-1}$ (filled symbols)
and  $\xi = 1.0$ kms$^{-1}$}
\end{figure}

\begin{table}[b]
\caption{Abundances and atmospheric parameters}
\begin{center}
\begin{tabular}{cccccccc}
\multispan8{\hrulefill}\\
\#&
$V_0$&
$T_{eff}$&
$\rm log~ g$&
$\rm [Fe/H]\hfill$&
$\rm [Mg/Fe]\hfill$&
$\rm [Fe/H]\hfill$&
$\rm [Mg/Fe]\hfill$\\
 & &  K & &$\xi=2$kms$^{-1}$ &
$\xi=2$kms$^{-1}$ &$\xi=1$kms$^{-1}$ &$\xi=1$kms$^{-1}$ 
\\
\multispan8{\hrulefill}\\
105 & 17.55 & 5041 & 2.59 & $-0.5$ & $+0.1$ & $-0.17 $ & $-0.11$\\
115 & 17.84 & 4953 & 2.36 & $-1.0$ & $+1.2$ & $-0.41 $ & $+0.61$\\
124 & 17.00 & 4891 & 2.21 & $-0.6$ & $+0.3$ & $-0.35 $ & $+0.02$\\
128 & 17.40 & 4778 & 1.93 & $+0.7$ & $-0.5$ & $+1.28 $ & $-0.50$\\
139 & 17.78 & 4891 & 2.21 & $-0.1$ & $+0.4$ & $+0.23 $ & $-0.50$\\
141 & 17.45 & 4977 & 2.42 & $-0.6$ & $+0.4$ & $-0.36 $ & $+0.12$\\
142 & 17.55 & 5118 & 2.81 & $-0.7$ & $+0.2$ & $-0.49 $ & $+0.14$\\
201 & 17.54 & 5003 & 2.49 & $+0.1$ & $-0.0$ & $+0.59 $ & $+0.17$\\
\multispan8{\hrulefill}\\
\end{tabular}
\end{center}
\end{table}

\section{Observations}

We used the multi-object-spectroscopy (MOS) mode
of the EMMI instrument on the NTT 3.5m telescope at ESO
La Silla. The resolving power 
was about 1500, the usable spectral range was from about 480 nm
to about 620 nm. 
We acquired spectra of 57 stars,
the log of the observations and further details
may be found in Marconi et al (1998b).

\section{Radial velocities and abundances}

The range 480-530 nm was used to determine heliocentric radial velocities
using cross-correlation with synthetic spectra
as templates 
We considered the stars with heliocentric radial velocities in the
range 100-180 kms$^{-1}$ to be members of Sagittarius, following
Ibata et al (1997), this left us with a sample of 23 stars.
In order to estimate abundances we 
defined six spectral indices which measure the Mgb triplet
and some Fe and iron-peak elements features.
We developed an iterative procedure 
which makes use of the SYNTHE code (Kurucz 1993)
to determine the abundances
which best match the observed to the synthetic indices;
details may be found in Marconi et al (1998b).
The procedure requires that the atmospheric parameters
T$_{\rm eff}$, log g and $\xi$ be fixed for each star.
The colour $(V-I)_0$ of Marconi et al (1998a)
was used to determine the effective temperatures
using the calibration of Alonso et al (1996); this
calibration refers to dwarfs only, however 
it is theoretically known that the $(V-I)$ colour depends
weakly on gravity; we expect the error 
introduced by neglecting the gravity dependence of $(V-I)$
to be on the order of 100 K. 
The isochrones of Straniero, Chieffi \& Limongi (1997) for 
an age of 8 Gyr and [Fe/H]$=-0.5$ were used to estimate
log g. Microturbulence cannot be determined from this intermediate-resolution
data, we therefore performed the computations for both 1 kms$^{-1}$
and 2 kms$^{-1}$, these values cover the range usually found in
cool giants and allow to estimate the effect of microturbulence
on the derived abundances.
This work is still in progress;
results  for 8 out of the 23 stars with
a radial velocity consistent with Sagittarius membership, are
 presented in Table 1
and displayed in Figure 1.

\section{Discussion}

Our stars have been selected in order to highlight 
the spread in metallicity derived from the colour-magnitude
diagram and in fact among the stars analyzed we see a spread
of over one dex. However the metallicities range from super-solar
to [Fe/H]$\approx -1.0$, about 0.5 dex more metal-rich than the range
implied by the photometry.
A systematic error in the spectroscopic or photometric
(or both !) abundance estimates could reconcile the two results.
However closer inspection of figure 1 reveals that of the 8 analyzed stars
3 are solar or super-solar while the remaining 5 show little or no
dispersion in metallicity. It is legitimate to ask whether the metal-rich
stars belong in fact to the Bulge, in spite of their radial velocity.
This hypothesis may not be ruled out, we note however that the [Mg/Fe]
ratios appear to be solar or sub-solar. 
The large errors associated with our data preclude any 
firm conclusion, however, taken at face value, the 
[Mg/Fe] in our metal-rich stars appears to be 
different from the enhanced
ratios displayed by the Bulge K
giants of McWilliam \& Rich (1994), figure 20a.
Our findings are in keeping with those of Smecker-Hane, McWilliam \& Ibata
(1998), who, on the basis of Keck HIRES spectra, found 2 out of 7 stars
to be metal-rich and with $\alpha$ elements O and Ca under-abundant with
respect to solar.

\end{document}